\documentstyle{article}
\newlength{\defaultparindent}
\newenvironment{Default Paragraph Font}{}{}
\input{tcilatex}

\begin{document}

\begin{center}
{\bf Localization of gravitational field energy and a procedure proposed}

{\bf for its experimental verification}

Miroslav Sukenik$^{a}$, Jozef Sima$^{a}$and Julius Vanko$^{b}$

$^{a}$Slovak Technical University, Dep. Inorg. Chem., Radlinskeho 9, 812 37
Bratislava, $^{b}$Comenius University, Dep. Nucl. Physics, Mlynska dolina F1,

842 48 Bratislava, Slovakia

e-mail: sima@chtf.stuba.sk; vanko@fmph.uniba.sk
\end{center}

Abstract

Introduction of Vaidya metrics into the Expansive Nondecelerative Universe
model allows to localize the energy of gravitational field. On the
assumption that there is an interaction of long-range gravitational and
electromagnetic fields, the localization might be verified experimentally.
In this contribution some details on such an experiment are given.

\begin{center}
Introduction
\end{center}

The law of universal gravitation was formulated by I. Newton in 1666. 250
years later A. Einstein showed that his equations describing gravitation had
wavelike solutions, similar to the wavelike solutions of Maxwell's
equations. Just as a time-varying charge distribution excites
electromagnetic waves, so a time-varying mass distribution will excite
gravitational waves. Yet, scientific community is still waiting for a direct
detection of gravitational waves. Moreover, contrary to the other
fundamental forces, there is neither comprehensive theory allowing to
quantify, localize and measure the gravitational field nor any experimental
evidence on the above mentioned characteristics of the gravitational field.
One of the main sources of this marking time lies in the Schwarzschild
metrics that has been generally accepted in elaborated models of the
Universe and gravitation. Usage of this metrics in models of the Universe
has led to a conclusion on the impossibility to localize the gravitational
field outside a body. The model of Expansive Nondecelerative Universe (ENU)
is virtually one of the first, if not the only, which is fully consistent
with the laws of nature (the requirement which must be satisfied by any
acceptable model), its predictions and conclusions are in accordance with
experimental observations and, thanks to the introduction of Vaidya metrics
into its mathematical tools, it provides, contrary to other models, a mode
of localization and quantification of the gravitational energy.

Within the research focused on unification of all four fundamental forces
several important theoretical results have emerged, no experimental data on
a mutual interaction and/or interference of any of them with gravitation
are, however, known. One of the principal obstacles lies in substantially
different range of the forces. From this viewpoint only the gravitational
and electromagnetic forces are similar since the both are far-reaching and
their intensity varies inversely with the square of the distance between two
bodies (masses or charges). These similarities have led us to considerations
on the theoretical and experimental treatment of possibilities to verify the
existence of interference of these forces. In this contribution theoretical
background of the verification is given.

\begin{center}
Theoretical Background
\end{center}

The density of gravitational field in the domain of weak fields is described
by Tolman's equation

$\epsilon _{g}=-\frac{R.c^{4}}{8\pi .G}\qquad \qquad \qquad \qquad \qquad
\qquad \qquad \qquad $(1)

where $R$ is the scalar curvature. In Schwarzschild metrics

$R$ = 0\qquad \qquad \qquad \qquad \qquad \qquad \qquad \qquad \qquad \qquad
(2)

that is interpreted as an impossibility to localize the gravitational energy
outside a body. In the Expansive Nondecelerative Universe model (further
ENU) [1 - 3] the gauge factor a is expressed as

$a=c.t_{c}=\frac{2G.M_{u}}{c^{2}}\qquad \qquad \qquad \qquad \qquad \qquad \
\ \ \ $(3)

where $M_{U}$ is the mass of the Universe and $t_{c}$ is the cosmologic time
and at present

$a\cong 10^{26}m$ \qquad \qquad \qquad \qquad \qquad \qquad \qquad \qquad\ \
\ (4)

$t_{c}\cong 4.5\times 10^{17}s$ \qquad \qquad \qquad \qquad \qquad \qquad
\qquad\ (5)

It follows from eq. (3) that in ENU the matter creation occours. The total
energy of the Universe must, however, be exactly zero [4]. It is achieved by
a simultaneous gravitational field creation, the energy of which is $E$ $<$
0. Such a Universe may, therefore, permanently expands with the velocity of
light $c$ and the fundamental conservation laws are still observed.

Due to the matter creation, Schwarzschild metrics may not be used and should
be replaced by another kind of metrics allowing to take the matter creation
phenomenon into account. Such a property exhibits Vaidya metrics [5]. Since
it holds in ENU [2, 3]

$\frac{dm}{dt}=\frac{m_{o}}{t_{c}}$ \qquad \qquad \qquad \qquad \qquad
\qquad \qquad \qquad\ \ \ \ \ (6)

where $m_{o}$ is the rest mass of a body, using Vaidya metrics and eq. (6),
relation (7) defining the scalar curvature is obtained

$R=\frac{3R_{g(m)}}{a.r^{2}}\qquad \qquad \qquad \qquad \qquad \qquad \qquad
\qquad $ (7)

in which $R_{g(m)}$ is the gravitational radius of a body with the mass $m$.
It follows [1] from (1) and (7) that

$\epsilon _{g}=-\frac{3mc^{2}}{4\pi ar^{2}}\qquad \qquad \qquad \qquad
\qquad \qquad \qquad \ \ \ \ \ $(8)

where $\epsilon _{g}$ is the density of gravitational field energy emitted
by a body with the mass $m$ in the distance $r$. Relation (8) can be
rewritten as

$\epsilon _{g}=\frac{3E_{g}}{4\pi .\lambda ^{3}}\qquad \qquad \qquad \qquad
\qquad \qquad \qquad \qquad $(9)

where $E_{g}$ is the energy of a gravitational quantum and $\lambda $ is its
Compton wavelength

$\lambda =\frac{h.c}{2\pi .E_{g}}\qquad \qquad \qquad \qquad \qquad \qquad
\qquad \qquad $(10)

Substituting in (9) for (10) and comparing the result with (8), the
expression for an energy quantum $E_{g}$ is obtained [1]:

$\left| {}\right. E_{g}\left. {}\right| =\left( \frac{m.h^{3}.c^{5}}{8\pi
^{3}a.r^{2}}\right) ^{1/4}$ \qquad \qquad \qquad \qquad\ \ \ (11)

where $E_{g}$ is the quantum of gravitational energy created by a body with
the mass $m$ in the distance $r$. Relation (11) is in conformity with the
limiting values: the maximum energy is represented by the Planck energy, the
minimum energy equals the energy of a photon with the wavelength identical
to the Universe dimension ($a$ =$\lambda $ ) .

Gravitational output $P_{g}$ , defined as the amount of gravitational energy
emitted by a body with the mass $m$ in a time unit is given as (12)

$P_{g}=\frac{d}{dt}\int \epsilon _{g}dV=-\frac{m.c^{3}}{a}=-\frac{m.c^{2}}{%
t_{c}}$ \qquad \qquad\ \ \ (12)

\begin{center}
Experimental verification
\end{center}

Both the gravitational and electromagnetic forces are of long-range nature
and their intensity decreases with the square of distance. Supposing an
interaction of the fields and stemming from the localization of the
gravitational field energy, an experiment may be proposed in which the
interference of the fields might lead to a change in the mass of the tested
body acting as an emitter. To perform the experiment, the following
requirements must be satisfied.

(a) A body - emitter with the mass $m_{e}$ should emit photons with the
energy $E_{ph}$ identical to the energy of Earth gravitational field quanta $%
E_{g,Ea}$ . Taking the known mass $m_{Ea}$ of the Earth (5.97 x 10$^{24}$
kg), its mean radius $r_{Ea}$ (6.37 x 10$^{6}$ m) and the values of the
other members in (11) into account, it follows

$E_{g,Ea}=\left( \frac{m_{Ea}.h^{3}.c^{5}}{8\pi ^{3}a.r_{Ea}^{2}}\right)
^{1/4}\cong 1.49eV$ \qquad \qquad \qquad (13)

that corresponds to the wavelength

$\lambda \cong 834nm$ \qquad \qquad \qquad \qquad \qquad \qquad \qquad
\qquad\ (14)

(this is a wavelength from the near-infrared region used in
telecommunications). One of the conditions, therefore, is that the emitter
should emit photons of the wavelength 834 nm. Such photons might interfere
with quanta of the gravitational field of the Earth.

(b) Relation (12) states that the electromagnetic output of the emitter $%
P_{e}$ is to be

$P_{e}=\frac{m_{e}.c^{2}}{t_{c}}$ \qquad \qquad \qquad \qquad \qquad \qquad
\qquad \qquad (15)

Relations (14) and (15) represent the conditions of equality of the
electromagnetic and gravitational output of the emitter. If, in principle,
may exist an interference of the electromagnetic and gravitational fields of
the emitter and the Earth, such an interference might happen and be observed.

(c) Since electronic transitions associated to excited state formation and
radiation deactivation are accompanied by vibrational-rotational (i.e.
thermal) relaxation, and the radiation output of the emitter is usually
influenced by its temperature, within the proposed experiment it will be
necessary to maintain constant temperature.

(d) After the required energy and output of the emitter $P_{e} $ are
reached, the emitter should be weighted at different directions of the
emission. The interference should manifest itself as a change in the emitter
weight that, given a balance with a sufficient sensitivity is used, should
be registered.

(e) There can be a necessity to adjust the wavelength and energy output
values (the accuracy of their values depends on the accuracy of all
parameters in used relations).

(f) Providing the radiation is not monochromatic (a common case) the output
of that with 834 nm (or that of adjusted wavelength) must correspond to that
required by relation (15).

(g) The mass $m_{e}$ of the emitter is its own mass not including peripheral
parts.

\begin{center}
Conclusions
\end{center}

1. The proposed experimental verification of interference of the
gravitational and electromagnetic fields may lead to successful results
providing three conditions are fulfiled: the ENU model describes the
Universe in a correct way; there is in principle a possibility of the
interference; conditions stated in paragraphs (a) to (g) are met.

2. Any changes in the mass of an emitter caused by the discussed
interference would be of far-reaching scientific and technological
importance.

3. The present contribution deals only with the theoretical aspects of the
experiment, its realization requires solutions of many practical tasks.

\begin{center}
References
\end{center}

1.J. Sima, M. Sukenik, Preprint: qr-qc/9903090

2. V. Skalsky, M. Sukenik, Astrophys. Space Sci., 178 (1991) 169

3. V. Skalsky, M. Sukenik, Astrophys. Space Sci., 181 (1991) 153

4. S. W. Hawking, A Brief History of Time, From the Big Bang to Black Holes,
Bantam Books, New York, 1988, p. 129

5. P.C. Vaidya: Proc. Indian Acad. Sci., A33 (1951) 264

\end{document}